\documentclass[preprint,showpacs,amssymb,aps,prd]{revtex4}

\begin{document}

%
\title{Black Hole Thermodynamics from 
\\
Near-Horizon Conformal Quantum Mechanics}

\author{Horacio E. Camblong$^{1}$
and
Carlos R. Ord\'{o}\~{n}ez$^{2,3}$}

\affiliation{
$^{1}$
Department of Physics, University of San Francisco, San
Francisco, California 94117-1080
\\
$^{2}$ Department of Physics, University of Houston, Houston,
Texas 77204-5506
\\ 
$^{3}$
World Laboratory Center for Pan-American Collaboration in Science and
Technology,
\\
University of Houston Center, Houston, Texas 77204-5506
}

\begin{abstract}
The thermodynamics 
of black holes is shown to be directly induced by their
near-horizon conformal invariance.
This behavior is exhibited
using a scalar field as a probe of the black hole gravitational background,
for a general class of metrics in D spacetime dimensions
(with $D \geq 4$).
The ensuing analysis is based on conformal quantum mechanics,
within a hierarchical near-horizon expansion.
In particular, 
the leading conformal behavior provides
the correct quantum statistical properties
for the Bekenstein-Hawking entropy,
with the near-horizon physics 
governing the thermodynamics from the outset.
Most importantly:
(i) 
this treatment 
reveals the emergence of
holographic properties;
(ii) 
the conformal coupling parameter
is shown to be related to the Hawking temperature;
and (iii) 
Schwarzschild-like coordinates,
despite their ``coordinate singularity,'' 
can be used
self-consistently to describe the thermodynamics
of black holes. 
\end{abstract}


\maketitle

\section{Introduction}
\label{sec:introduction}

The Bekenstein-Hawking entropy $S_{\rm BH}$~\cite{Bekenstein_entropy},
the Hawking temperature $T_{H}$, and the Hawking 
effect~\cite{Hawking_temperature}
are well-established features of black hole 
thermodynamics~\cite{BH_thermo_reviews} whose universality
points to the existence of a quantum gravitational theory.
Moreover, the statistical-mechanical derivations
of the entropy $S_{\rm BH}$ from string theory~\cite{BH_string}
and loop quantum gravity~\cite{BH_loop}
verify that these results do not depend
on the details of the underlying 
quantum theory of gravity.
In addition, the thermodynamic properties appear to originate
from the event horizon~\cite{padmanabhan_review_04},
within two major categories:
(i) 
those arising from the relationship
$S_{\rm BH} = {\mathcal A}/4$
between the entropy 
and the horizon area ${\mathcal A}$;
(ii)
those related to the
near-horizon conformal symmetry.
The first category has led to 
't Hooft's brick-wall model~\cite{thooft:85} and
the thermal-atmosphere proposal~\cite{thorne_zurek}---which
suggest an origin of the entropy from within 
a ``Planck-length skin'' of the 
horizon~\cite{thooft:85}---and subsequently to
 the holographic principle~\cite{holographic} and
the AdS/CFT correspondence~\cite{AdS/CFT}.
In the second category, the neighborhood of the horizon
displays a peculiar SO(2,1) 
conformal symmetry~\cite{wyb:74,jackiw,fubini_beyond,conformal_geometry};
this kind of black-hole near-horizon 
invariance~\cite{strominger:98,carlip:near_horizon,solodukhin:99}
has been generalized to its supersymmetric 
extensions~\cite{BH_superconformal},
and related to horizon states~\cite{gov:BH_states,gupta:BH},
to the thermodynamics~\cite{gupta:BH},
and to the Calogero model~\cite{gupta:calogero,calogero_black_holes}.
Moreover, in Refs.~\cite{carlip:near_horizon,solodukhin:99},
the thermodynamics is explicitly connected with 
the underlying near-horizon conformal field theory
through the Cardy formula.

With these ideas in mind,
in this paper we develop a framework 
within which black hole thermodynamics emerges
from the near-horizon conformal symmetry
as the central guiding principle. Furthermore,
we display  a {\em direct and explicit connection between 
the conformal symmetry and the thermodynamics\/}:
(i)
the Hawking temperature is determined from near-horizon
consistency requirements and traced to the conformal symmetry;
(ii)
the Bekenstein-Hawking entropy can
be interpreted within a brick-wall model 
through a near-horizon conformal quantum mechanics;
(iii) 
the determination of 
the entropy as a physical observable 
leads to a natural cutoff of the order of the Planck length.
Hence, our work supports the concept
that the quantum degrees of
freedom of a black hole appear to reside
on its horizon and should arise
from a Planck-scale quantum theory of gravity.

In this paper we adopt the
metric conventions of Ref.~\cite{misner_thorne_wheeler}
and choose natural units
$\hbar =1$, $c=1$, and $k_{B}=1$; 
by contrast, the $D$-dimensional gravitational
constant $ G_{N}^{(D)} $  is displayed in appropriate expressions, 
especially in Sec.~\ref{sec:geometric_renormalization}.
 In Sec.~\ref{sec:QFT_probe}
we consider a scalar field in the gravitational background
and study its near-horizon behavior.
In Sec.~\ref{sec:thermodynamics}
we develop
the general framework for the computation of thermodynamic properties.
In Sec.~\ref{sec:geometric_renormalization}
we provide a renormalization of the entropy in a geometric manner, 
which we implement with the aid of 't Hooft's brick-wall model. Finally,
these ideas are critically reexamined in Sec.~\ref{sec:conclusions}.

\section{Field Modes and Near-Horizon Expansion}
\label{sec:QFT_probe}

The conjecture that {\em the horizon encodes the quantum 
properties of a black hole\/}~\cite{thooft:85} can be tested by considering 
a quantum field as a probe of the gravitational background. 
This method has been extensively used 
in the literature dating back to the early seminal
works of the 1970s, including Ref.~\cite{Hawking_temperature}.
The main purpose of our paper is to apply this well-known
technique to show that the near-horizon conformal 
symmetry of Refs.~\cite{jackiw,gov:BH_states,gupta:BH}
governs the leading thermodynamics
of the Bekenstein-Hawking entropy and the Hawking temperature.
These properties can be seen most easily
for the particular case of
an action ($D \geq 4$)
\begin{equation}
S
=
-
\frac{1}{2}
\int
d^{D} x
\,
\sqrt{-g}
\,
\left[
g^{\mu \nu}
\,
\nabla_{\mu} \Phi
\, 
\nabla_{\nu} \Phi
+ 
m^{2} \Phi^{2}
+  \xi R \Phi^{2}
\right]
\; ,
\label{eq:scalar_action}
\end{equation}
which describes the coupling of a scalar field $\Phi$ 
to the background metric $g_{\mu \nu}$
through its covariant derivatives
$\nabla_{\mu} \Phi$
and to the curvature scalar $R$.
In addition, 
we assume a metric
\begin{equation}
 ds^{2}
=
- f (r) \,  dt^{2}
+
\left[ f(r) \right]^{-1} \, dr^{2}
+ r^{2} \,
 d \Omega^{2}_{(D-2)}
\; 
\label{eq:RN_metric}
\end{equation}
(where
$
d \Omega^{2}_{(D-2)}
$ is the metric 
on the unit sphere $S^{D-2}$),
which includes the Reissner-Nordstr\"{o}m geometries in $D$ spacetime 
dimensions~\cite{mye:86}, extensions with a cosmological 
constant, and related stringy black-hole solutions with additional 
charges~\cite{ortin}. Equation~(\ref{eq:scalar_action}) can be 
generalized to include additional fields; this possibility,
which may lead to ambiguities~\cite{thooft:85,frolov},
is critically revisited in Sec.~\ref{sec:conclusions}.

The  expansion of the quantum field
$\Phi$ in generalized Schwarzschild coordinates $(t, r,\Omega )$,
\begin{equation}
 \Phi(t, r, \Omega )
= \sum_{n,l, m }
\left[
 	a_{nl m }
\phi_{nl m } (r, \Omega )
\,
e^{-i\omega_{nl}t}
 	+ a^{\dagger}_{nl m }
\phi^{*}_{nl m } (r, \Omega )
\,
e^{i\omega_{nl}t}
 	\right]
\; ,
\label{eq:field_Fourier_expansion}
\end{equation}
involves creation and annihilation operators 
subject to the usual canonical commutation relations and 
a complete set of orthonormal modes
$
\phi_{nl m } (r,\Omega )
= 
Y_{l m } \big(  \Omega \big)
\chi (r)
u_{nl} (r)
$
that satisfy the equation
$
\left[ 
\Box 
- 
\left(
m^{2} 
+ 
\xi R
\right)
 \right] \phi \, e^{\mp i \omega t}
=
0
$;
the discrete index
$n$ corresponds to enclosing the system in a spherical box
for the thermodynamic analysis.
For a metric~(\ref{eq:RN_metric}),
the angular dependence of the modes is
 given by the ultraspherical harmonics
$Y_{l m } \big( \Omega \big)$~\cite{ultraspherical},
with eigenvalues  $\lambda_{l,D} 
=l(l+D-3) $,
while the choice
$
\chi (r)
=
[f(r)]^{-1/2} 
\,
r^{-(D-2)/2}
$
generates a Liouville transformation~\cite{forsyth:Liouville}
that reduces the equation for the radial 
part to its normal form
\begin{equation}
u_{nl}''(r) 
+
\mathfrak{I} (r; \omega_{nl}, \alpha_{l,D}  ) 
\,
u_{nl}(r) 
=0
\;  
\label{eq:Klein_Gordon_normal_radial}
\end{equation}
for every  particular frequency $\omega_{nl}$.
Thus,
the reduction of the 
field~(\ref{eq:field_Fourier_expansion})
to its normal modes 
induces an {\em effective quantum mechanics\/}.
With
 $f \equiv f(r)$ and 
the parameters
$\alpha_{l,D} 
=
\lambda_{l,D}  + \nu^{2}
=
\left[
l + (D-3)/2 
\right]^{2}
$ 
and
$
\nu = (D-3)/2
$,
the {\em effective interaction\/}
$\mathfrak{I}   $
is given by
\begin{equation}
\mathfrak{I}    
(r; \omega, \alpha_{l,D}  ) 
=
\frac{ 1 }{f^{2}}
\left[
 \omega^{2} + 
\frac{f'^{2}}{4}
\right]
-
\frac{1}{ f }
\frac{\alpha_{l,D} }{ r^{2} } 
-
\frac{1}{ f }
\left(  m^{2} + \xi R \right)
+ R_{rr} 
+
\left[
\left(
\frac{1}{ f } - 1
\right)
\nu^{2} 
+ \frac{1}{4}
\right]
\frac{1}{r^{2}} 
\;  ,
\label{eq:Klein_Gordon_normal_radial_invariant}
\end{equation}
where
$
R_{rr}
=
-
f''/2f
-
(D-2) f'/(2rf)
$
is the radial component of the Ricci tensor.
The effective interaction $\mathfrak{I}$ includes
two noteworthy terms:
the first one,  leading to the 
SO(2,1) conformal interaction in Schwarzschild coordinates; and
the second one, which gives the only dependence
of $\mathfrak{I}  (r; \omega, \alpha_{l,D}  ) $
with respect to 
the field angular momentum.

The near-horizon conformal symmetry 
can be studied 
by considering an expansion of Eq.~(\ref{eq:Klein_Gordon_normal_radial}),
with the variable 
$x= r -r_{+}$,
where $r_{+}$ is the root of the 
equation $f(r)=0$ defining the outer event horizon ${\mathcal H}$.
In this paper, we will consider
the {\em nonextremal\/} case, with
$
f'_{+} 
\equiv f'(r_{+}) 
\neq 0
$
(the extremal case
is known to involve a number of subtleties~\cite{frolov}).
Consequently, the terms in 
Eq.~(\ref{eq:Klein_Gordon_normal_radial_invariant}) can be 
reduced with 
$f''/f  
\stackrel{(\mathcal H)}{\sim}
f''_{+}/(f'_{+}x) $ and 
$f'/f 
\stackrel{(\mathcal H)}{\sim}
 1/x$, together with 
$r \stackrel{(\mathcal H)}{\sim}
r_{+}$,
where $\stackrel{(\mathcal H)}{\sim}$
stands for the hierarchical expansion about ${\mathcal H}$;
then, the {\em leading\/} terms,
of order $O(1/x^{2})$, become asymptotically dominant
and Eq.~(\ref{eq:Klein_Gordon_normal_radial}) turns into
\begin{equation}
u''(x)
+
\frac{ \lambda }{x^{2}}
\,
\left[ 1 + O(x) \right]
u (x)
\stackrel{(\mathcal H)}{\sim}
0
\;  ,
\label{eq:Klein_Gordon_conformal}
\end{equation}  
which 
is driven by the interaction
$
V_{\rm eff} (x) = 
- 
\lambda/x^{2}
$,
with a one-dimensional effective Hamiltonian
$H = {p}_{x}^{2}
-  \lambda /x^{2}$.
In Eq.~(\ref{eq:Klein_Gordon_conformal}),
by abuse of notation:
 $u(r) \equiv u(x)$, and
\begin{equation}
\lambda = 
 \Theta^{2} +
\frac{1}{4} 
\, , \; \; \; \; 
\Theta
= 
\frac{\omega}{  f'_{+} } 
\; .
\label{eq:conformal_interaction_coupling}
\end{equation}  
The corresponding physics,
known as conformal quantum 
mechanics~\cite{camblong,renormalization_CQM},
is invariant under general  ``effective-time 
(${\mathcal T}$)
reparametrizations,'' 
where ${\mathcal T}$
is the variable conjugate to the Hamiltonian $H$.
These transformations involve~\cite{jackiw}  translations 
generated  by $H$, scalings due to the
 dilation operator
$
D
\equiv
{\mathcal T} 
H
-
\left( 
p_{x} x + x p_{x} 
\right)/4$,
and
 translations of reciprocal ${\mathcal T} $
due to the special conformal operator
$
 K
\equiv
2 {\mathcal T}  D -
{\mathcal T}^{2}H
+
x^{2}/4
$.
The commutators
\begin{equation}
[D,H]
= - i \hbar H
 \;  ,
\; \;
[K,H]
= - 2 i \hbar D
\;  ,
\; \;
[D, K]
=  i \hbar K
\;  ,
\label{eq:naive_commutators}
\end{equation}
define a noncompact
SO(2,1) $\approx$ SL(2,R) Lie algebra~\cite{wyb:74},
which summarizes the near-horizon
dynamics of the field
in Schwarzschild coordinates.
While the relevance of this symmetry 
for black hole thermodynamics was
first discussed in Refs.~\cite{gov:BH_states,gupta:BH},
the full-fledged form of the conformal 
coupling~(\ref{eq:conformal_interaction_coupling})
for arbitrary frequencies $\omega$ has not been properly
recognized.
In contrast to the work of
Refs.~\cite{gov:BH_states,gupta:BH},
we show herein that
this frequency dependence is a crucial ingredient
for the Hawking temperature $T_{H}$
and the Bekenstein-Hawking entropy $S_{\rm BH}$.
Specifically:
(i)
Eq.~(\ref{eq:conformal_interaction_coupling})
describes an effective system with
the conformal symmetry algebra~(\ref{eq:naive_commutators})
in the strong-coupling 
regime ($\lambda> 1/4$);
(ii)
such system
experiences the characteristic pathologies of 
{\em singular quantum mechanics\/},
which, as we will see in the next section,
lead to a {\em divergent contribution
to the density of modes that governs the thermodynamics\/}.

The apparent simplicity of Eq.~(\ref{eq:Klein_Gordon_conformal})
has completely erased all information about 
the additional dynamical degrees of freedom of the field:
the angular-momentum variables. 
For the calculation of the entropy
we need a generalized expansion that includes the leading order with respect
to angular momentum.
As this dynamical dependence appears in only one term,
$ \alpha_{l,D}/(f r^{2}) $, 
in Eq.~(\ref{eq:Klein_Gordon_normal_radial_invariant}),
the leading orders become
\begin{equation}
\mathfrak{I}  (r; \omega, \alpha_{l,D}  ) 
=
\left\{
\left[
\frac{ \omega^{2} }{(f'_{+})^{2}}
+
 \frac{1}{4} 
\right]
x^{-2}
-
\frac{ \alpha_{l,D}  }{f_{+}' r_{+}^{2}}
\frac{1}{x}
\right\}
\, 
\left[ 1 + O(x) \right]
\; . 
\label{eq:scalar_field_in_BH_background_5_invariant_NH}
\end{equation}
In the hierarchical
 expansion~(\ref{eq:scalar_field_in_BH_background_5_invariant_NH}),
one can see the reason for the necessity 
to keep track of this additional angular-momentum
dependence. 
While all other terms in 
Eq.~(\ref{eq:Klein_Gordon_normal_radial_invariant})
become negligible for sufficiently small $x$,
the term
$ \alpha_{l,D}/(f r^{2}) $ 
can become comparable to 
the leading order $x^{-2}$ in 
Eq.~(\ref{eq:scalar_field_in_BH_background_5_invariant_NH}),
for sufficiently high values of 
$ \alpha_{l,D} $.
In other words,
for sufficiently high angular momentum $l$, the 
near-horizon expansion needs to be supplemented by an
angular-momentum contribution of order  $x^{-1}$.
This additional term provides 
a cutoff that carries the necessary phase-space
information for the statistical counting of degrees of freedom.
Thus, 
{\em it is the interplay between 
the conformally invariant near-horizon leading term 
and the field angular momentum
that completely determines the thermodynamics\/}; in 
 Sec.~\ref{sec:geometric_renormalization}, we will see that this
competition leads directly 
to the holographic property $S_{\rm BH} = {\mathcal A}/4$.

\section{Thermodynamics and Spectral Functions}
\label{sec:thermodynamics}

The central concept behind 
the statistical mechanics of the field $\Phi$ 
is the existence of {\em thermal averages\/}.
For the static spacetimes with metrics~(\ref{eq:RN_metric}), 
thermodynamic equilibrium at temperature $T= 1/\beta$ can be 
established from the periodicity of the Euclidean time $\tau_{E} = -i t$
in finite-temperature field theory.
In the seminal work of Ref.~\cite{conical_GH},
the Hawking temperature $T=T_{H}$ was shown to be 
the unique value required for the removal of a bolt singularity 
of the near-horizon Euclidean metric.
In terms of the conformal parameter $\Theta$:
\begin{equation}
T_{H}= 
\frac{  f'_{+}}{4 \pi} 
=
\left( 4 \pi \, \frac{\Theta}{\omega} \right)^{-1}
\; 
\label{eq:Hawking_temperature}
\end{equation}
follows from the near-horizon expansion of
the $(\tau_{E},r)$ sector of the metric, which takes the 
two-dimensional polar-coordinate form
$
f(r) \, d\tau_{E}^{2} + [f(r)]^{-1} d r^{2}
\stackrel{(\cal H)}{\sim}
\rho^{2} d \alpha^{2} + d \rho^{2}
$, with
$f^{-1} dr^{2} = d \rho^{2}$~\cite{zee}.
This argument unambiguously shows that
{\em the thermodynamics is dictated by the near-horizon
conformal physics.\/}
However, a complete characterization of thermal equilibrium
entails self-consistency within conformal quantum mechanics; 
in principle, for every frequency $\omega$ and $T_{H}$ given in
Eq.~(\ref{eq:Hawking_temperature}), this amounts to the realization 
of thermal equilibrium through a Boltzmann 
factor~\cite{Hawking_temperature,gibbons_perry}
$\exp \left[ - \omega/T_{H} \right]$, 
as in the complex-path method of 
Refs.~\cite{padmanabhan_HawkingT,padmanabhan_HawkingT_extra,vagenas}.
Incidentally,
the invariance of the temperature and surface gravity
of a stationary black hole
under conformal transformations of the metric,
$g_{\mu \nu} \rightarrow \Omega^{2} g_{\mu \nu}$,
is a well-known property~\cite{jacobson_conf_H-temp};
however, the connection between the approach of
Ref.~\cite{jacobson_conf_H-temp} 
and that of our paper---based on the symmetry 
algebra~(\ref{eq:naive_commutators})---is 
not immediately obvious.
These issues will be considered in a forthcoming 
publication, using the SO(2,1) conformal 
interaction.

With the temperature~(\ref{eq:Hawking_temperature})
in the canonical ensemble, the thermodynamic functions
can be computed in the usual way~\cite{thooft:85,frolov};
for example, starting with the free energy $F$ and density operator
$\rho = e^{- \beta (:H:-F)}$,
the entropy
$
 S  \equiv  - 
{\rm Tr}\left[ 
\rho \ln \rho
\right]
 	= 
\beta^{2}
\partial F/\partial\beta$
is given by
\begin{equation}
S 
=
- \int_{0}^{\infty}
d \omega 
\,
\ln (1 - e^{-\beta \omega})
\,
\left[
\left(
\omega \frac{d}{d \omega} + 2 
\right) 
\frac{dN(\omega)}{d\omega}
\right]
\; ,
\label{eq:entropy_formula_2}
\end{equation}
which follows from the familiar expression for 
a free field~\cite{thooft:85,frolov}
through integration by parts.
In Eq.~(\ref{eq:entropy_formula_2}), the nontrivial effects of the 
spacetime curvature are carried by the spectral function $N(\omega) $,
which measures the cumulative number of modes
associated with the field equation~(\ref{eq:Klein_Gordon_normal_radial}).
In turn, the mode ordering  $\{nl  m \}$ is governed by
 Sturm's theorem~\cite{ince:sturm} for a given 
effective potential~(\ref{eq:Klein_Gordon_normal_radial_invariant}),
so that
${\mathcal N}_{l} ( \omega )
= {\mathcal Z}_{l} ( \omega ) +1 $
and $ {\mathcal Z}_{l} ( \omega ) $
are the ordinal number and number of zeros of the eigenfunction
$u_{nl}(r)$ in Eq.~(\ref{eq:Klein_Gordon_normal_radial})
for every value of $\omega$.
As a result,
\begin{equation}
N (\omega) 
=
\sum_{
\stackrel{n,l,  m }{
\mathfrak{I}  (r; \, \omega_{nl}, \alpha_{l,D}  ) 
\leq
\mathfrak{I}  (r; \, \omega, \alpha_{l,D}  ) }
}
\, 
1
\,
=
\sum_{ l }
\,
g_{l}
\,
{\mathcal N}_{l}
( \omega )
\; ,
\label{eq:Sturm_cumulative_number_of_states}
\end{equation}
where
$g_{l}=
(2l+D-3) (l+ D-4)!/[l! (D-3)!]
$
is the multiplicity
of $Y_{l m } \big( \Omega \big)$~\cite{ultraspherical}.

The spectral function  $N (\omega) $
can be computed with the algorithm of  
Eq.~(\ref{eq:Sturm_cumulative_number_of_states})
combined with the semiclassical approximation~\cite{thooft:85,frolov},
which involves a linear combination of
\begin{equation}
 u_{\pm} (r) 
=
\left[
k_{\alpha_{l,D} } (r)
\right]^{-1/2} 
 \exp \left[
\pm i \int^{r} 
k_{\alpha_{l,D}} (r') dr'
\right]
\; ,
\label{eq:WKB_wave_functions}
\end{equation}
i.e., the familiar WKB wave functions
with a local wavenumber
$ k_{\alpha_{l,D}} (r)$.
For the relevant domain, namely, in the neighborhood of the horizon,
a Langer-corrected wavenumber~\cite{langer} 
\begin{equation}
 k_{\alpha_{l,D}} (r)
=
k_{\alpha_{l,D}} (r_{+}+ x )
=
\sqrt{ 
\mathfrak{I}   (r_{+}+ x; \omega , \alpha_{l,D}  ) 
- \frac{1}{ 4 x^{2} }}
\;
\label{eq:WKB_wavenumber}
 \end{equation}
is required to deal properly with the coordinate 
singularity.
The ordinal number
\begin{equation}
{\mathcal N}_{l}
( \omega )
=
\int_{\mathcal I} 
\,
 	k_{ \alpha_{l,D }} (r)
dr
\; 
\label{eq:semiclassical_ordinal_number}
\end{equation}
is obtained from the wave functions~(\ref{eq:WKB_wave_functions}),
with an integration range
${\mathcal I} $ in the spatial region outside the horizon,
limited by the semiclassical restriction within the turning points.
In addition, the nontrivial angular-momentum sum 
in Eq.~(\ref{eq:Sturm_cumulative_number_of_states})
can be approximated in the semiclassical regime
by means of the rule~\cite{semiclassical_ang_momentum_sum}
\begin{equation}
\sum_{l} g_{l} \,
F(\alpha_{l}) 
\sim 
\frac{1}{   \Gamma (D-2) }
\int_{0}^{ \infty} 
d \alpha \,
\alpha^{D/2- 2} 
\, 
F(\alpha)
\;  .
\label{eq:semiclassical_angular_momentum_sum}
\end{equation}
As a result,
substituting Eqs.~(\ref{eq:semiclassical_ordinal_number})
and (\ref{eq:semiclassical_angular_momentum_sum})
in Eq.~(\ref{eq:Sturm_cumulative_number_of_states}),
we obtain 
\begin{equation}
 N(\omega) = 
\frac{1}{\pi \, \Gamma (D-2) }
\int_{0}^{ \infty}  
d \alpha \,
\alpha^{D/2- 2} 
\,
\int_{\mathcal I} 
dr
\,
 	k_{ \alpha} (r)
\; ,
\label{eq:semiclassical_number_of_states}
\end{equation}
where the semiclassical interval ${\mathcal I}$ is limited by a right
turning point $r_{\rm max} = r_{\rm max} (\alpha)$, which is defined by the
zero of the radicand in Eq. (14). Reciprocally, if the order of integration
is reversed, an angular momentum cutoff $\alpha_{\rm max}$ can be defined
for a given $x$;
this is implicitly given by
$
\mathfrak{I}   (r_{+}+ x; \omega , \alpha_{\rm max}  ) 
= 1/ (4 x^{2})
$.

As it stands,
Eq.~(\ref{eq:semiclassical_number_of_states}) 
describes the physics of the scalar field
in the gravitational background,
including the effects associated with all relevant scales.
In particular, it contains: (i) 
its ordinary bulk behavior; (ii)
effects of the near-horizon physics, which correspond to 
the sector $r \sim r_{+}$; (iii)
additional terms arising from the intermediate region. 
For the relevant near-horizon physics,
a systematic near-horizon expansion can be applied
to Eq.~(\ref{eq:semiclassical_number_of_states}) 
and then transferred to all relevant thermodynamic
quantities.
The leading orders 
in Eq.~(\ref{eq:scalar_field_in_BH_background_5_invariant_NH}) 
call for the use of the Langer prescription~(\ref{eq:WKB_wavenumber}),
which yields the replacement
$\lambda/x^{2}
 \rightarrow
\lambda/x^{2}
-
1/4x^{2}
=
\Theta^{2}/x^{2} 
$.
Therefore, 
from Eqs.~(\ref{eq:conformal_interaction_coupling}), 
(\ref{eq:scalar_field_in_BH_background_5_invariant_NH}),
and
(\ref{eq:WKB_wavenumber}), 
\begin{equation}
 k_{\alpha_{l,D}} 
=
 k_{\alpha_{l,D}} 
(r= r_{+}+x; \Theta, 
\alpha_{l,D}
)
\stackrel{(\mathcal H)}{\sim}
\sqrt{ \frac{\Theta^{2} }{x^{2}} 
\, [ 1 + O(x) ] 
- \frac{ A( r_{+})
\,
\alpha_{l,D} } {{x} }
\, [ 1 + O(x) ] \;
}
\; ,
\label{eq:WKB_wavenumber_nh}
\end{equation}
where $A( r_{+})= 
1/ (f'_{+} \,  r_{+}^{2} )$
 stands for the angular-momentum coefficient.

The leading orders of the corresponding spectral 
function~(\ref{eq:semiclassical_number_of_states}) become
\begin{equation}
 N(\omega) 
\stackrel{(\mathcal H)}{\sim}
\frac{ \Theta}{\pi \, \Gamma (D-2) }
\int_{0}^{ \infty} 
d \alpha \,
\alpha^{D/2- 2} 
 \, 
\int_{a}^{ x_{\rm max} (\alpha) } 
\frac{dx}{x}
\sqrt{ 1 
- 
\frac{ A( r_{+}) 
\,
\alpha }{ \Theta^{2}
    }
\, x }
\;
 [ 1 + O(x) ] 
\; ,
\label{eq:WKB_number_of_states}
\end{equation}
where $x_{\rm max} = r_{\rm max} - r_{+} $ and $a$ is a coordinate
cutoff.

Two important conclusions stem from 
this analysis, 
from Eq.~(\ref{eq:WKB_number_of_states}):
\begin{itemize}
\item
The conformal interaction involves 
an effective ``coupling parameter'' $\Theta^{2}$ rather than $\lambda$.
This parameter emerges
from the near-horizon physics alone  
and provides a 
 {\em conformal wave number\/} 
$
k_{\rm conf} (x)
= \Theta/x 
$
in Eq.~(\ref{eq:WKB_wavenumber_nh}).

\item
The angular-momentum coefficient
$A( r_{+})$,
needed for the mode counting~(\ref{eq:WKB_number_of_states}),
is due to the $S^{D-2}$ foliation of the metric
and yields an angular-momentum degeneracy factor
$
\chi_{\alpha_{l,D}} (x)
= 
\sqrt{ 1 
- 
A( r_{+}) 
\alpha_{l,D}
x /\Theta^{2}   } $
that modifies the 
$k_{\rm conf} (x)$ 
in Eq.~(\ref{eq:WKB_wavenumber_nh}).
\end{itemize}

Finally, a simple rescaling of the integral with respect 
to $\alpha$ shows that
\begin{equation}
 N(\omega) 
\stackrel{(\mathcal H)}{\propto}
\Theta^{D-1} 
\,
\left[ A ( r_{+}) \right]^{ -(D-2)/2 }
\,
\lim_{a \rightarrow 0}
\int_{a}^{x_{1}} \frac{dx}{x^{D/2}}
\;
 [ 1 + O(x) ] 
\; ,
\label{eq:WKB_number_of_states_2}
\end{equation}
where $a$ is a near-horizon
coordinate cutoff for the radial variable $r$
and $x_{1}$ is an arbitrary upper limit.
Unfortunately, two major flaws of
Eq.~(\ref{eq:WKB_number_of_states_2}) prevent
a meaningful application of this formula.
 First, the integral in Eq.~(\ref{eq:WKB_number_of_states_2}) 
is divergent with respect to the limit  $a \rightarrow 0$,
and this singular behavior is transferred to
all thermodynamic functions, including the 
entropy~(\ref{eq:entropy_formula_2}).
This {\em ``ultraviolet catastrophe''\/}~\cite{BH_thermo_reviews}, which
can be viewed as due to the divergent near-horizon redshifts,
signals the existence of new 
quantum gravitational physics near the horizon
and requires an appropriate {\em regularization\/}
of the theory.
One of the novel features of the approach presented herein
is the description of this 
``ultraviolet catastrophe''
in {\em Schwarzschild coordinates\/},
as directly arising from singular {\em conformal quantum mechanics\/}.
Second, the naive use
of a radial cutoff $a$ as a finite adjustable parameter
cannot work as this is merely a coordinate assignment;
instead, the thermodynamic functions should be 
recast in terms of physical observables---a 
{\em renormalization\/} of the theory.
In conclusion, there is a way of treating the divergence and the 
noncovariant nature of $a$ simultaneously:
the concurrent use of real-space 
renormalization and a geometric redefinition of $a$.
In particular, the brick-wall model~\cite{thooft:85}
provides an implementation 
of this regularization. 
This is the problem to which we now turn.

\section{Geometric Renormalization}
\label{sec:geometric_renormalization}

The divergent behavior 
of the spectral function~(\ref{eq:WKB_number_of_states_2}) 
and of the associated thermodynamics
has a simple physical interpretation.
The framework 
defined by a field action~(\ref{eq:scalar_action}) in a gravitational 
background~(\ref{eq:RN_metric})
is but an {\em effective theory\/} that calls for 
modifications in the ultraviolet sector, 
as the event horizon is approached.
In a generic sense, this is the ansatz known as 
 't Hooft's ``brick-wall model,'' 
according to which the relevant part of the entropy $S$
in Eq.~(\ref{eq:entropy_formula_2})
 arises from a ``thermal atmosphere'' extending a 
few Planck lengths above the horizon,
and whose ultimate origin lies in a full-fledged 
quantum theory of gravitation.

In our approach,
the ultraviolet cutoff $a$ in 
Eq.~(\ref{eq:WKB_number_of_states_2}) provides 
an approximate coordinate value
leading to a scale for the transition to more fundamental 
short-distance physics.
As such, $a$ is a particular value of the
Schwarzschild coordinate $r$ rather than a proper length scale.
For the {\em geometrization\/} of the theory,
what is needed is
 a {\em proper distance\/}~\cite{thooft:85} 
\begin{equation}
\rho (x)
= 
\left[
\ell_{P}^{(D)}
\right]^{-1}
\,
\int_{ r_{+}}^{r_{+}+x} 
| g_{rr}(r)|^{1/2} dr
\stackrel{(\mathcal H)}{\sim}
\frac{2}{ \ell_{P}^{(D)} \sqrt{ f'_{+} } }
\,
\sqrt{x}
\;
 [ 1 + O(x) ] 
\; 
\label{eq:geometrical_radial_distance_n-h}
\end{equation}
from the horizon,
which we write in dimensionless
form  with respect to the $D$-dimensional
Planck length 
$\ell_{P}^{(D)} =
\left[ G_{N}^{(D)} \right]^{1/(D-2)}$.
In particular, the
proper ``geometrical elevation'' $h_{D} $ 
of the ``brick wall''
(away from the horizon)
 can be identified as $h_{D}  = \rho (a)$.
In a more restricted sense,
the regularization of the theory
can be implemented by enforcing a boundary condition 
at the location defined by the 
 coordinate parameter $a$.
In particular, a sharp cutoff
in the integral of 
Eq.~(\ref{eq:WKB_number_of_states_2})
is equivalent to
the use of a Dirichlet boundary condition 
\begin{equation}
 \Phi(t, r = a ,\Omega ) = 0
\label{eq:Dirichlet}
\; ;
\end{equation}
this assignment is
 a consequence of the selection of a 
semiclassical left turning point.
However, the existence of fairly general results in conformal quantum 
mechanics, which are independent of the selection of the 
ultraviolet physics~\cite{camblong,renormalization_CQM},
suggests that different boundary conditions 
are likely to yield the same physics.

The redefinition involved in Eq.~(\ref{eq:geometrical_radial_distance_n-h})
permits the geometrization of 
Eq.~(\ref{eq:WKB_number_of_states}),
\begin{equation}
 N(\omega)
\stackrel{(\mathcal H)}{\sim}
\frac{2 \Theta}{\pi}
\,
\int_{ h_{D}  }
\frac{ d \rho}{\rho}
\;
\varrho_{D} 
\biggl(
{\alpha_{\rm max}} (\rho)  
\biggr)
\; ,
\label{eq:WKB_number_of_states_3}
\end{equation}
where the angular-momentum degeneracy is described by the
 weight function
\begin{eqnarray}
\varrho_{D} \left(     {\alpha_{\rm max}}   \right)
& = &
\frac{1}{   \Gamma (D-2) }
\int_{0}^{\alpha_{\rm max}} 
d \alpha \,
\alpha^{D/2- 2} \; 
\sqrt{ 1 - \frac{\alpha}{ \alpha_{\rm max} } }
\label{eq:ang_momentum_degeneracy_weight}
\\
& \stackrel{(\mathcal H)}{\sim} &
{\mathcal C}_{D}
\,
 \frac{
 \hat{{\mathcal A}}_{D-2}  }{4}
\,
\left(
\frac{\Theta}{\rho}
\right)^{D-2}
\; ,
\label{eq:ang_momentum_degeneracy_weight_explicit}
\end{eqnarray}
with
$
 \hat{{\mathcal A}}_{D-2} 
= 
\Omega_{(D-2)} 
\,
\left[
r_{+}/\ell_{P}^{(D)}
\right]^{D-2}
$
being the $(D-2)$-dimensional horizon area in Planck units,
given in terms of  $\Omega_{(D-2)} = 
2 \pi^{(D-1)/2}/\Gamma
\left(
(D-1)/2
\right)
$.
In Eqs.~(\ref{eq:WKB_number_of_states_3})
and (\ref{eq:ang_momentum_degeneracy_weight_explicit})
and hereafter, 
the higher-order terms
of the near-horizon expansion are omitted;
beta-function identities give the
numerical constant 
$
{\mathcal C}_{D}
= 
2^{D} \, 
\Gamma \left( D/2  \right)/
\pi^{D/2- 1} \; \Gamma (D) 
$; and
the angular-momentum cutoff,
from Eqs.~(\ref{eq:WKB_number_of_states})
and (\ref{eq:geometrical_radial_distance_n-h}),
becomes
$
{\alpha_{\rm max}} (\rho)
=
4 \,
\left[ r_{+}/\ell_{P}^{(D)}  \right]^{2}
\,
\Theta^{2}/\rho^{2} 
$.

A number of remarks are in order.
Equation~(\ref{eq:WKB_number_of_states_3})
shows the interplay between the weight 
function~(\ref{eq:ang_momentum_degeneracy_weight})
and the purely conformal contribution
$
 N_{\rm CQM}(\omega)
=
2 \Theta/\pi
\,
\int_{ h_{D}  }
d \rho/\rho
$,
which would otherwise lead to a renormalized conformal logarithmic
counting of states~\cite{renormalization_CQM}. In contrast to
this logarithmic behavior, in the case of black hole thermodynamics,
the angular-momentum degeneracy weight changes the
distance scaling in
Eq.~(\ref{eq:WKB_number_of_states_3}),
due to the 
additional dependence implicit through
the ``cutoff''
$
{\alpha_{\rm max}} (\rho)
$.
Furthermore,
Eq.~(\ref{eq:ang_momentum_degeneracy_weight_explicit})
shows the presence of two distinct
contributions,
in addition to the numerical constant
 ${\mathcal C}_{D}$:
the ``holographic factor'' 
$
 \hat{{\mathcal A}}_{D-2} /4$
and the 
factor associated with the
``conformal part'' of the angular-momentum cutoff,
$
(\Theta/\rho)^{D-2}
$.
For the class of metrics considered in this work, 
the holographic factor emerges from a  phase-space
contribution that can be traced to the
horizon hypersurface.
In turn,
the ``conformal part'' of the angular-momentum cutoff factor
is due to the competing effects of the conformal interaction, 
parametrized via the effective coupling $\Theta^{2}$,
and the angular-momentum term.
Correspondingly, from
Eqs.~(\ref{eq:WKB_number_of_states_3})--(\ref{eq:ang_momentum_degeneracy_weight_explicit}),
\begin{equation}
 N(\omega)
\stackrel{(\mathcal H)}{\sim}
{\mathcal N}_{D}
\,
 \frac{ \hat{{\mathcal A}}_{D-2}  }{4}
\,
\left[
\Theta (\omega)
\right]^{D-1}
\; ,
\label{eq:WKB_number_of_states_4}
\end{equation}
where 
${\mathcal N}_{D}=  \left\{ 2 {\mathcal C}_{D}/[ (D-2) \, \pi ] \right\}
\left[ h_{D} \right]^{-(D-2)} $
is a numerical constant arising from
phase-space counting of modes and
from measuring the cutoff elevation
 $ h_{D} $.
Most importantly,
Eq.~(\ref{eq:WKB_number_of_states_4})
shows that the angular momentum 
contributes to the horizon degrees of freedom
through
$\varrho_{D} \left(     {\alpha_{\rm max}}   \right)$,
while
the conformal interaction mainly induces the degrees
of freedom due to radial displacements and
associated with the SO(2,1) symmetry.

The {\em geometric renormalization\/} of the spectral functions, 
leading to Eq.~(\ref{eq:WKB_number_of_states_4}), 
transfers to all thermodynamic  quantities. In particular, this procedure 
should apply to the entropy~(\ref{eq:entropy_formula_2}).
The fundamental concept already displayed by 
Eq.~(\ref{eq:WKB_number_of_states_4}),
is that the entropy is a surface contribution induced by the horizon.
Our derivation displays this 
$(D-2)$-dimensional feature 
in its most transparent form 
as arising from the summation over angular-momentum  
degrees of freedom.
Correspondingly,
this also suggests the property known as holography,
whose realization for black-hole entropy
appears to be related to the conformal nature of the
near-horizon expansion.
Specifically, substituting
Eq.~(\ref{eq:WKB_number_of_states_4})
in Eq.~(\ref{eq:entropy_formula_2}),
\begin{equation}
S
\stackrel{(\mathcal H)}{\sim}
{\mathcal S}_{D}
\,
\left(
\frac{ 4 \pi}{ \beta
f_{+}'}
\right)^{D-1}
\,
S_{\rm BH}
\; ,
\label{eq:entropy_brick_wall}
\end{equation}
where the expected Bekenstein-Hawking entropy is
\begin{equation}
S_{\rm BH}
= 
\frac{1 }{4}
\,
\hat{{\mathcal A}}_{D-2}  
\label{eq:Bekenstein_Hawking_entropy}
\; ,
\end{equation}
 and the numerical constant
\begin{equation}
{\mathcal S}_{D}
= 
\frac{D (D-1)}{2^{D-1}}
\,
 {\mathcal N}_{D} 
\,
{\mathcal J}_{D} 
=
\left[
\frac{ \pi^{1-3D/2} }{2^{D-2}}
\, D
\zeta (D) \Gamma (D/2-1) 
\right]
 \left[ h_{D} \right]^{-(D-2)}
\; 
\label{eq:elevation_factor_in_entropy}
\end{equation}
has been evaluated 
in terms of the Riemann zeta function $\zeta (z)$
from the integral
\begin{equation}
{\mathcal J}_{D} = - \int_{0}^{\infty}
d \eta \,
\eta^{D-2} \ln \left(1- e^{-2\pi \eta} \right)
=
\frac{\zeta (D) \, \Gamma (D-1)}{ (2\pi)^{D-1}}
\; .
\end{equation}

Finally, the entropy~(\ref{eq:entropy_brick_wall}) reduces to the
expected holographic result~(\ref{eq:Bekenstein_Hawking_entropy}),
but only after two additional identifications are made.
First,
the factor
$\left[ 4 \pi/( \beta
f_{+}') \right]^{D-1}$
can be set equal to unity, due to the Hawking-temperature 
assignment~(\ref{eq:Hawking_temperature}).
The second identification involves 
the  factor~(\ref{eq:elevation_factor_in_entropy}),
which should be set equal to unity;  this condition 
determines the ``elevation'' 
\begin{equation}
 h_{D} 
=
\frac{1}{2}
\left[
 D \zeta (D) \Gamma (D/2-1) \pi^{1-3D/2}
\right]^{1/(D-2)}
\; ,
\label{eq:brick_wall_geometrical_elevation}
\end{equation}
of the brick wall above the horizon.
For example,  for $D=4$~\cite{thooft:85}, 
 $ h_{D} $ in
Eq.~(\ref{eq:brick_wall_geometrical_elevation})
 reduces to $1/\sqrt{90 \pi}$.
Thus, when physical units are
restored in terms of the Planck length $\ell_{P}^{(D)}$,
this distance becomes $ H_{D} = 
h_{D}  \; \ell_{P}^{(D)}$, whose order 
of magnitude is comparable to that of
$\ell_{P}^{(D)}$.

In conclusion,
the entropy~(\ref{eq:Bekenstein_Hawking_entropy}) follows
quite naturally within conformal quantum mechanics
and requires a real-space regulator whose concomitant 
invariant distance is of the order of the
Planck length. 
Moreover, our derivation shows two important features:
(i) {\em the entropy is a $(D-2)$-dimensional property\/}
induced by the near-horizon expansion and implemented through
the angular-momentum phase-space counting of states;
(ii) {\em the temperature is purely conformal\/}.
These universal properties are driven by the near-horizon
symmetry and apply to a large class
of black holes and any number of dimensions, thus
suggesting the existence  of an underlying 
order arising from the Planck scale.

\section{Conclusions}
\label{sec:conclusions}

In this paper we have considered the near-horizon conformal 
symmetry of a broad class of black-hole metrics and described the 
emergence of thermodynamic behavior induced by the existence 
of an event horizon.
Specifically, we have rederived the Hawking 
temperature~(\ref{eq:Hawking_temperature})
and Bekenstein-Hawking entropy~(\ref{eq:Bekenstein_Hawking_entropy})
almost exclusively from this conformal symmetry.
In the case of the entropy,
an appropriate treatment of 
the angular momentum degrees of freedom directly
relates to the horizon area,
with the conformal sector requiring an
effective-field-theory type of renormalization 
such as that within the brick-wall model.
The ensuing  
symmetry-based characterization of the thermal nature of black holes
ascribes the singular behavior of thermodynamic quantities
to the physics within a ``Planck-length skin''
surrounding the horizon. In addition, our work:

(i) Provides strong additional evidence that the physical origin of 
the quantum-mechanical degrees of freedom of a black hole 
can be traced to within a Planck scale of the event horizon. 

(ii) Shows the need for new physics 
near the Planck scale, manifested through the existence of
an invariant radial distance from the horizon
where the theory breaks down~\cite{thooft:horizons}.

 (iii)
May prove useful in identifying the relevant parts of quantum gravity that 
are responsible for the thermodynamic behavior of black holes.

A number of critical remarks are in order,
as the brick wall model poses several puzzling questions.
First, the scalar field-action of Eq.~(\ref{eq:scalar_action}) 
can be extended to involve any number of ``species,'' 
with different types of fields; this ambiguity implies
a possible dependence on the number and type of 
species~\cite{thooft:85,BH_thermo_reviews,frolov}.
Second, when this generalized
action is applied to the computation of the
entropy as in Eq.~(\ref{eq:entropy_brick_wall}), the identification 
of the Bekenstein-Hawking result~(\ref{eq:Bekenstein_Hawking_entropy})
with a numerical prefactor of $1/4$ requires a fine-tuning 
of the cutoff~\cite{thooft:85,BH_thermo_reviews,frolov},
as in Eq.~(\ref{eq:brick_wall_geometrical_elevation}).
In other words, the entropy prefactor is not calculable in this approach,
thus being subject to renormalization;
however, the result
still has two remarkable features that suggest
its possible correctness: the area dependence and the expected 
 order of magnitude.
Finally, it was first shown in Ref.~\cite{susskind_uglum} and 
subsequently confirmed
in other papers~\cite{renormalization_Newton_confirmed}
that the brick-wall contribution to the entropy
can be interpreted as being absorbed
by a renormalization of Newton's gravitational constant $G_{N}$.
A possible interpretation
of these multiple ambiguities is that
the species dependence of the entropy prefactor
is compensated by a corresponding dependence 
of the renormalization of $G_{N}$;
this is confirmed by miscellaneous renormalization
approaches~\cite{frolov,renormalization_BW_miscellaneous,BH_confirmed_Boulware}.
Notwithstanding any unresolved issues,
the results of our paper appear
to be extremely {\em robust\/} and
confirm the relevance of the {\em conformal
aspects of black hole thermodynamics\/}; in particular,
the temperature and the Hawking effect are independent 
of any particular regularization model. 
In this regard, the near-horizon conformal 
symmetry appears to be central to black hole thermodynamics, even though its
physical interpretation and relationship to spacetime symmetries 
of quantum gravity still remain elusive. In this context, 
it would be useful to uncover the meaning of our construction within 
an approach based on conformal field theories, as in the  work of 
Refs.~\cite{carlip:near_horizon,solodukhin:99}.
\acknowledgments{This research was supported 
by the National Science Foundation under Grant 
No.\ 0308300 (H.E.C.) 
and under Grant No.\ 0308435 (C.R.O.), and
by the University of San Francisco Faculty Development Fund
(H.E.C.).
We also thank 
Professor Cliff Burgess
for stimulating discussions and
Dr.\ Stanley Nel for generous travel support 
that facilitated the conception of this project.
}

\end{document}